# Infrared spectra of the water-CO₂ complex in the 4.3-3.6 μm region and determination of the ground state tunneling splitting for HDO-CO₂

T. Gartner,[1] C. Lauzin,[2] A.R.W. McKellar,[3] and N. Moazzen-Ahmadi[1,*]

[1] *Department of Physics and Astronomy, University of Calgary, 2500 University Drive North West, Calgary, Alberta T2N 1N4, Canada*

[2] *Institute of Condensed Matter and Nanosciences, Université catholique de Louvain, B-1348, Louvain-la-Neuve, Belgium*

[3] *National Research Council of Canada, Ottawa, Ontario K1A 0R6, Canada*

**Abstract**

Spectra of water - CO₂ dimers are studied using a tunable mid-infrared source to probe a pulsed slit jet supersonic expansion. $H_2O$-$CO_2$ and $D_2O$-$CO_2$ are observed in the CO₂ $\nu_3$ fundamental region ($\approx 2350$ cm$^{-1}$), $D_2O$-$CO_2$ is also observed in the $D_2O$ $\nu_3$ fundamental region ($\approx 2790$ cm$^{-1}$), and HDO-CO₂ is observed in the HDO O-D stretch fundamental region ($\approx 2720$ cm$^{-1}$), all for the first time in these regions. Analysis of the spectra yields excited state rotational parameters and vibrational shifts. They also yield the first experimental values of the ground state internal rotation tunneling splittings for $D_2O$-$CO_2$ (0.003 cm$^{-1}$) and HDO-CO₂ (0.0234 cm$^{-1}$). The latter value is a direct determination made possible by the reduced symmetry of HDO-CO₂. These results provide stringent and easily interpreted tests for theoretical water - CO₂ potential energy surface calculations.

*Corresponding author
Email address: nmoazzen@ucalgary.ca
Phone: 1-403-220-5394

# 1.    Introduction

Water and carbon dioxide molecules exist in abundance on our planet and are responsible for a plethora of natural phenomena that govern the world around us. Living organisms depend on both water and carbon dioxide for direct sustenance and respiration and both molecules are important for innumerable industrial processes that define our daily life. Our atmosphere is made up of anywhere from 0-4% water vapor and roughly 0.04% carbon dioxide, although these quantities are constantly in flux due to natural and manmade processes. Understanding and quantification of the possible interactions between these two molecules has diverse applications which include environmental chemistry, atmospheric science (both planetary and extrasolar) and the potential sequestration of $CO_2$ in water clathrates, among others.[1] A basic starting point of such studies is the multidimensional $H_2O$-$CO_2$ potential energy surface, though of course as cluster size grows many-body effects are also important. Probably the most stringent and direct tests for theoretical potential surfaces come from high resolution spectroscopy of water-$CO_2$ dimers, including various isotopes and excited vibrations which help to probe the surface in different regions.

High resolution water-$CO_2$ spectra were first reported in 1984 by Peterson and Klemperer,[2] who observed microwave pure rotational transitions involving $H_2O$-, $D_2O$-, and HDO-$CO_2$ and concluded that the dimer has a planar side-by-side structure with the O-H bonds pointing away from the $CO_2$, giving $C_{2v}$ point group symmetry as shown in Fig. 1. The intermolecular bond length (C atom to water O atom) was determined to be about 2.8 Å. There was evidence of hindered internal rotation of the water relative to the $CO_2$ around the symmetry axis. The magnitude of this internal rotation barrier was estimated to be "high", namely about 300 cm$^{-1}$. Subsequently, in 1998, Columberg et al.[3] considerably extended the original water-$CO_2$ microwave observations. Meanwhile, in 1992, an infrared spectrum of $H_2O$-$CO_2$ was observed by Block et al.[4] in the region of the $\nu_3$ (asymmetric stretch) fundamental band of $H_2O$ ($\approx$3750 cm$^{-1}$). Notably, their infrared analysis yielded an experimental value of 0.268 cm$^{-1}$ for the sum of the $H_2O$-$CO_2$ internal rotation tunneling splittings in the ground and excited ($H_2O$ $\nu_3$) states, from which the ground state value could be estimated as about 0.134 cm$^{-1}$. More recently, infrared spectra of $H_2O$-$CO_2$ and $D_2O$-$CO_2$ were observed[5] in the region of the water bending vibration (1595 and 1178 cm$^{-1}$, respectively), and spectra of $H_2O$-$CO_2$ were observed[6,7] in the 2OH overtone region, corresponding to the $(v_1, v_2, v_3) = (2,0,0)$ and $(1,0,1)$ vibrations of $H_2O$ ($\approx 7200$ cm$^{-1}$).

There have been a few theoretical *ab initio* calculations of the full (5D or higher) water-$CO_2$ potential energy surface, most recently those by Makarewitz,[8] by Wheatley and Harvey,[9] and by Wang and Bowman.[1] The latter surface has been used by Felker and Bačić [10] for detailed calculations of $H_2O$- and $D_2O$-$CO_2$ tunneling, rotational, and intermolecular vibrational energies. The calculated depth of the potential surface minimum ($D_e$) is around 1050 cm$^{-1}$, and the $H_2O$-$CO_2$ dissociation energy ($D_0$) is about 770 cm$^{-1}$.

In the present paper, the study of water-$CO_2$ infrared spectra is extended to include the $CO_2$ $v_3$ region ($\approx 2350$ cm$^{-1}$) for $H_2O$-$CO_2$ and $D_2O$-$CO_2$, the $D_2O$ $v_3$ region ($\approx 2790$ cm$^{-1}$) for $D_2O$-$CO_2$, and the O-D stretch ($\approx 2720$ cm$^{-1}$) for HDO-$CO_2$. We thereby obtain the first experimental values for the internal rotation tunneling splittings of $D_2O$-$CO_2$ ($\approx 0.003$ cm$^{-1}$) and HDO-$CO_2$ (0.02344 cm$^{-1}$). The latter value is a direct ground state measurement thanks to the reduced symmetry of HDO-$CO_2$, which enables our detection of weak *a*-type transitions in its spectrum in addition to the dominant *b*-type ones.

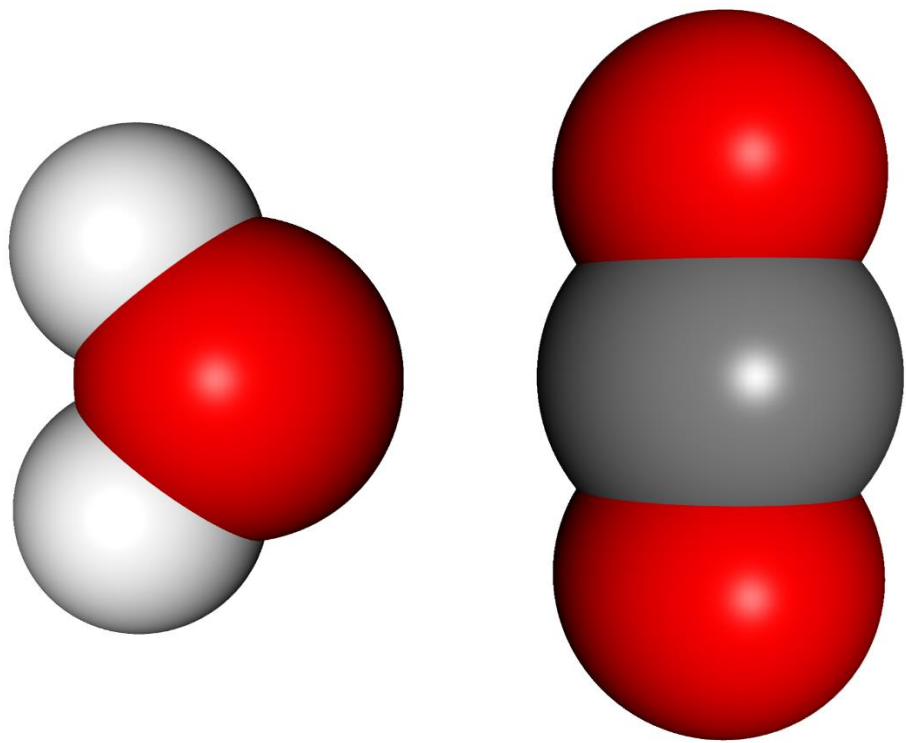

Figure 1. Equilibrium structure of the $H_2O$-$CO_2$ dimer. The intermolecular distance (C atom to $H_2O$ O atom) is about 2.8 Å.

## 2.    Methods

Spectra were recorded as described previously[11,12] using a pulsed supersonic slit jet expansion probed by a rapid-scan tunable infrared optical parametric oscillator source. The supersonic expansion mixtures contained about 0.01% water vapor ($H_2O$, $D_2O$, or $H_2O + D_2O$) plus 0.02 – 0.4% $CO_2$ in helium carrier gas with a backing pressure of about 10 atmospheres. The smaller $CO_2$ concentration was used for the spectra in the $CO_2$ $\nu_3$ region, and the larger concentration in the O-D stretch region. Wavenumber calibration was carried out by simultaneously recording signals from a fixed etalon and a room temperature reference gas cell. Spectral assignment, simulation, and fitting were made with the help of the PGOPHER software.[13]

## 3.    Results

### 3.1. $H_2O$-$CO_2$

The molecular symmetry group for the internally rotating water-$CO_2$ is $G_8$.[14] Due to the combined effects of the equivalent O nuclei in $CO_2$ and the equivalent H nuclei in $H_2O$, only $K_a$ = even rotational levels are allowed in the ground vibrational state of $H_2O$-$CO_2$ for the lower internal rotation tunneling state, which has a nuclear spin weight of 1, and only $K_a$ = odd levels for the upper tunneling state, which has a spin weight of 3. As mentioned above, the tunneling splitting is known to be approximately 0.134 cm$^{-1}$.[4] Our observed spectrum of $H_2O$-$CO_2$ in the region of the $CO_2$ $\nu_3$ fundamental is shown in the upper panel of Fig. 2. This band has $b$-type selection rules, with $\Delta K_a = \pm 1$. In the upper vibrational state, $K_a$ = odd levels are allowed for the lower tunneling state (spin weight 1), and $K_a$ = even levels for the upper tunneling state (spin weight 3). Rotational assignments in the spectrum were relatively straightforward thanks to the known ground state parameters,[3] and we assigned a total of 93 lines in terms of 97 transitions (there were four blended lines) with values of $K_a' = 0$ to 5 and $J' = 0$ to 9.

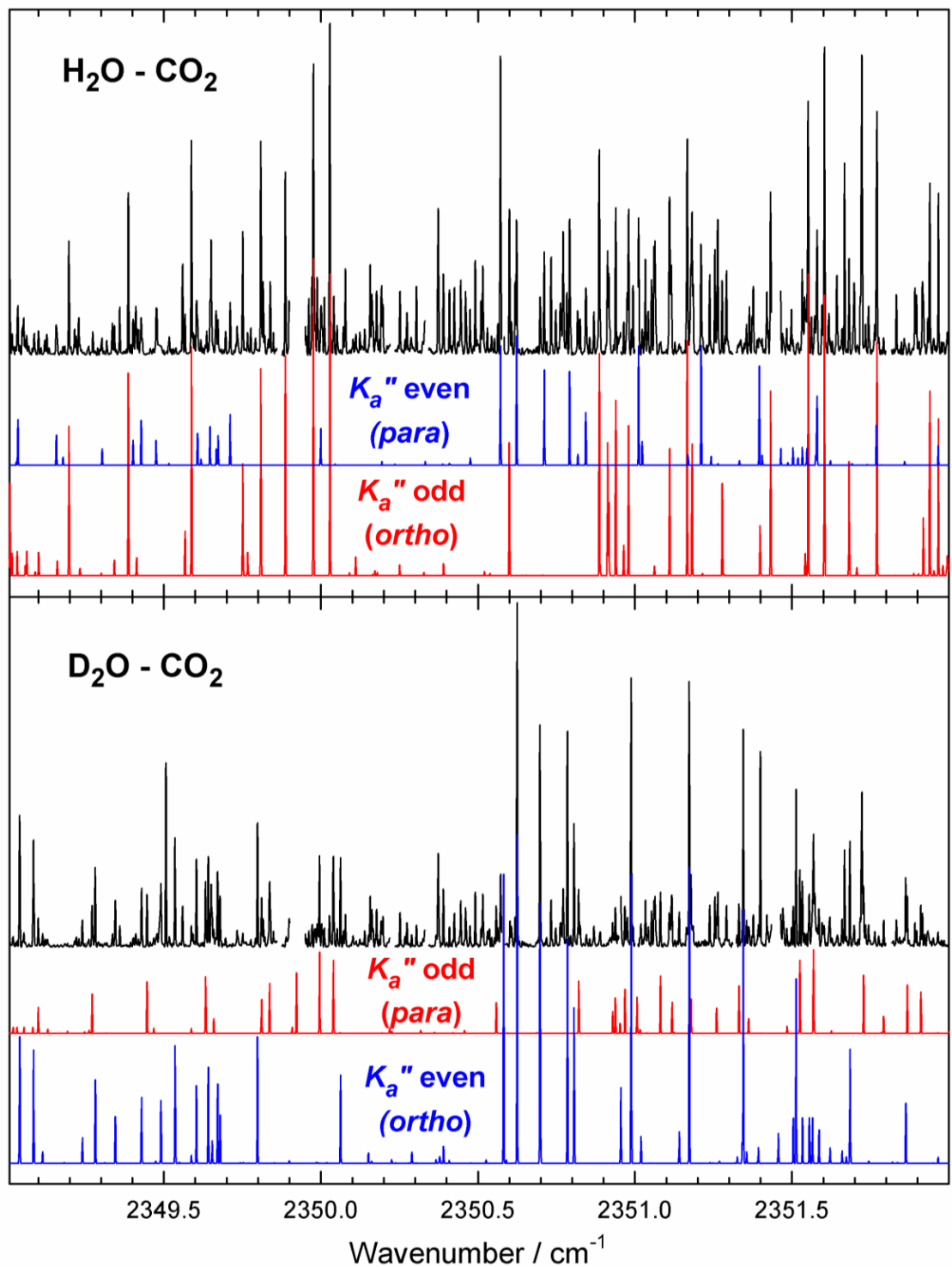

Figure 2. Observed and simulated spectra of H₂O- and D₂O-CO₂ in the region of the CO₂ ν₃ fundamental. Small gaps in the observed spectra correspond to regions of strong absorption by CO₂ and CO₂-He. The absorption lines in the observed spectrum not accounted for in the simulation are due to (CO₂)₂.

In the analysis, the ground state parameters were fixed at the values determined by Columberg et al.,[3] which were found to represent our observed ground state combination differences perfectly. (The revised ground state parameters of Bogomolov et al.[7] gave virtually identical results). For the excited vibrational state, the band origins and rotational parameters were varied separately for the lower and upper tunneling states, while the centrifugal distortion parameters were fixed at their ground state values.[3] Results are shown in Table 1, and these parameters were used for the simulated spectra in Fig. 2. The root mean square (rms) error for the 93 observed lines was 0.00021 cm$^{-1}$, essentially similar to the experimental precision. Since the transitions take place from lower to lower and upper to upper states, the small difference between the two band origins in Table 1 (0.0016 cm$^{-1}$) represents the difference between the tunneling splittings in the ground and excited ($CO_2$ $\nu_3$) states.

Table 1. Molecular parameters for $CO_2$-$H_2O$ (in cm$^{-1}$). [a]

| | $K_a{'}$ odd $K_a{''}$ even | $K_a{'}$ even $K_a{''}$ odd |
|---|---|---|
| $\nu_0$ | 2350.30095(7) | 2350.30255(5) |
| $A{'}$ | 0.3811376(61) | 0.3806881(57) |
| $B{'}$ | 0.1559383(69) | 0.1559413(33) |
| $C{'}$ | 0.1099601(59) | 0.1099791(26) |
| $A{''}$ | 0.3840898 | 0.3836397 |
| $B{''}$ | 0.155921644 | 0.155943835 |
| $C{''}$ | 0.110221172 | 0.110230183 |
| $\Delta_K{''}$ | -1.067 e-5 | |
| $\Delta_{JK}{''}$ | 1.11490 e-5 | |
| $\Delta_J{''}$ | 9.4692 e-7 | |
| $\delta_K{''}$ | 7.778 e-6 | |
| $\delta_J{''}$ | 2.8216 e-7 | |

[a] Quantities in parentheses are 1σ from the least-squares fit, in units of the last quoted digit. Ground state parameters are from Columberg et al.[3] Excited state centrifugal distortion parameters were fixed at the indicated ground state values.

### 3.2. D₂O-CO₂

For the $D_2O$-$CO_2$ ground vibrational state, $K_a$ = even levels are allowed for the lower tunneling state and $K_a$ = odd levels for the upper tunneling state, just as for $H_2O$-$CO_2$. But now the spin weights are 6 and 3, respectively, rather than 1 and 3. The $D_2O$-$CO_2$ tunneling splitting has been theoretically predicted[10] to be 0.0039 cm$^{-1}$, much smaller than that of $H_2O$-$CO_2$. We observed $D_2O$-$CO_2$ spectra in two separate regions, corresponding to the $\nu_3$ fundamental bands of $CO_2$ and $D_2O$.

The $D_2O$-$CO_2$ spectrum in the $CO_2$ region is shown in the lower panel of Fig. 2. It is actually very similar to that of $H_2O$-$CO_2$, with only a very small change in band origin and relatively small changes in rotational constants. The main difference is due to the spin weights, which cause even $K_a''$ transitions to be stronger in $D_2O$-$CO_2$ and weaker in $H_2O$-$CO_2$, while odd $K_a''$ transitions are the other way around. We assigned a total of 107 lines in terms of 117 transitions (ten blended lines) with values of $K_a'$ = 0 to 5, and $J'$ = 0 to 9. The $D_2O$-$CO_2$ spectrum in the $D_2O$ region is shown in Fig. 3. Here we assigned a total of 93 lines in terms of 98 transitions (five blended lines) with values of $K_a'$ = 0 to 4 and $J'$ = 0 to 8. This is also a $b$-type band since the $D_2O$ $\nu_3$ antisymmetric stretch dipole transition moment lies along the $D_2O$-$CO_2$ $b$-axis, so its general appearance is similar to that in the $CO_2$ region.

We checked a number of observed $D_2O$-$CO_2$ infrared ground state combination differences against those calculated using the parameters of Columberg et al.,[3] and found that the calculated values for the interval $K_a''$ = 3 − 5 were systematically too low by approximately 0.001 cm$^{-1}$, and those for $K_a''$ = 2 − 4 low by 0.0003 cm$^{-1}$. We therefore reanalyzed the ground state using our combination differences plus the microwave[2,3] data, giving a high relative weight (1200) to the latter to reflect the high microwave precision. The resulting ground state parameters are given in the lower part of Table 2. They agree with those of Columberg et al.[3] within the mutual error limits, but their precision is improved, especially for the parameters $A$ and $\Delta_K$. These ground state parameters were then held fixed for the analysis of the infrared spectrum, with results as given in the upper part of Table 2. The rms errors in the infrared fit were 0.00015 cm$^{-1}$ for the $CO_2$ region and 0.00040 cm$^{-1}$ for the $D_2O$ region. The larger residuals in the $D_2O$ region are a reflection of small but recognizable upper state perturbations (for example, the levels $J_{Ka,Kc}$ = $3_{1,3}$ and $5_{3,3}$ were each observed to be about 0.001 cm$^{-1}$ higher than calculated). There is almost no difference between the two $D_2O$-$CO_2$ band origins in the $CO_2$ region (as for $H_2O$-$CO_2$ above), but there is a significant difference in the $D_2O$ region. This is because the transitions in the latter case are from the lower to upper, and upper to lower, tunneling states, so that the band origin difference is the sum of the tunneling splittings in the ground and excited states. Tunneling splittings are discussed in more detail below (Sec. 4.1).

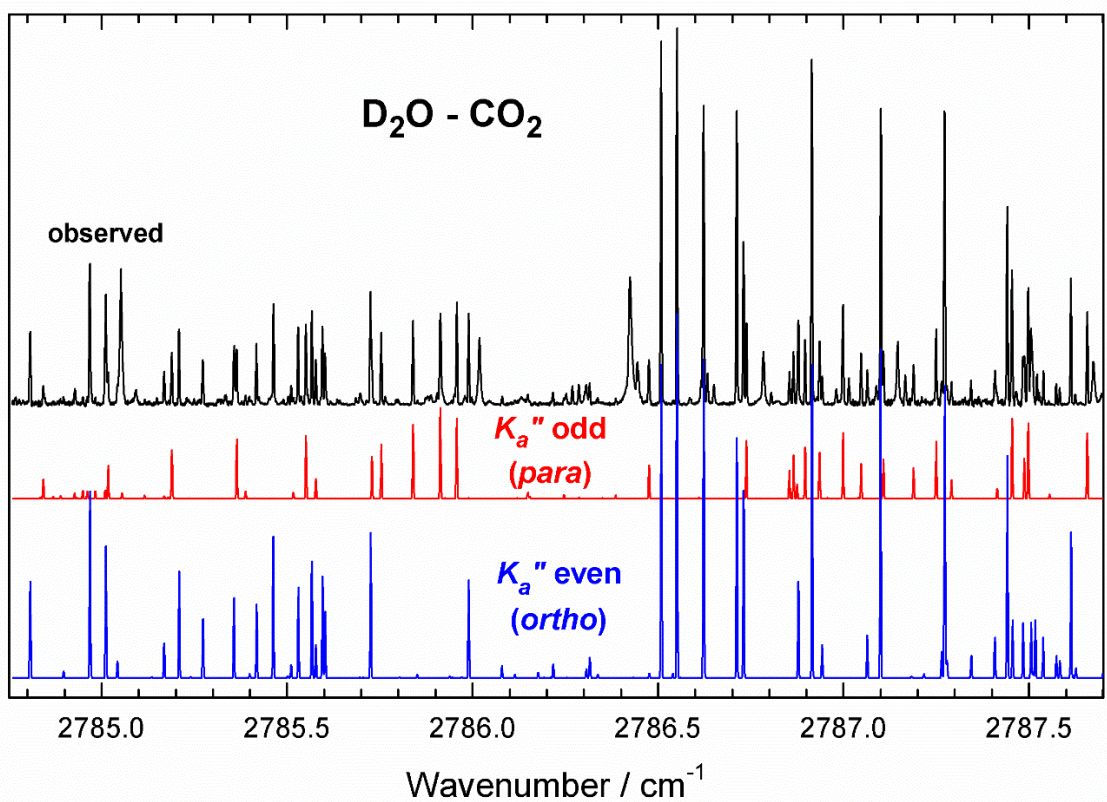

Figure 3. Observed and simulated spectra of $D_2O$-$CO_2$ in the region of the $D_2O$ $\nu_3$ fundamental

Table 2. Molecular parameters for $CO_2$-$D_2O$ (in cm$^{-1}$).[a]

| | $K_a'$ odd $K_a''$ even | | $K_a'$ even $K_a''$ odd | |
|---|---|---|---|---|
| $\nu_0$ | 2350.31240(7) | 2786.2364(1) | 2350.31259(8) | 2786.2306(1) |
| $A'$ | 0.3714975(70) | 0.374404(39) | 0.3714506(88) | 0.374265(44) |
| $B'$ | 0.1406504(43) | 0.1404079(87) | 0.1406555(84) | 0.140367(16) |
| $C'$ | 0.1015046(29) | 0.101603(10) | 0.1014894(54) | 0.101585(10) |
| $\Delta_K'$ | [-0.965e-5] | 1.50(30)e-5 | [-0.965e-5] | 1.50e-5 |
| $\Delta_{JK}'$ | [1.03530e-5] | -1.44(10)e-5 | [1.03530e-5] | -1.44 e-5 |
| $\Delta_J'$ | [7.356e-7] | 1.36(14)e-6 | [7.356e-7] | 1.36 e-6 |
| $A''$ | 0.3742939(12) | | 0.3742670(82) | |
| $B''$ | 0.14064516(29) | | 0.14064591(15) | |
| $C''$ | 0.10171025(27) | | 0.10171023(15) | |
| $\Delta_K''$ | -0.965(29)e-5 | | | |
| $\Delta_{JK}''$ | 1.03530(95)e-5 | | | |
| $\Delta_J''$ | 7.356(24)e-7 | | | |
| $\delta_K''$ | 7.062(67)e-6 | | | |
| $\delta_J''$ | 2.1118(58)e-7 | | | |

[a] Quantities in parentheses are 1$\sigma$ from the least-squares fit, in units of the last quoted digit. Ground state parameters are from our fit to previous microwave data[2,3] plus selected ground state combination differences from the present spectra. Excited state centrifugal distortion parameters for the 2350 cm$^{-1}$ band were fixed at ground state values, and those for the 2786 cm$^{-1}$ band were constrained equal for the two tunneling components, as indicated. Excited state $\delta_K$ and $\delta_J$ parameters were fixed at their ground state values.

### 3.3. HDO-$CO_2$

Internal rotation tunneling occurs in HDO-$CO_2$, just as for $H_2O$- and $D_2O$-$CO_2$, and we thus continue to fit the lower (even $K_a''$) and upper (odd $K_a''$) tunneling components with separate parameter sets. But now there is no distinction between *para* and *ortho* nuclear spin modifications and transitions between tunneling components are possible. We observed the HDO-$CO_2$ infrared spectrum in the region of the O-D stretch fundamental, as shown in Fig. 4 and assigned numerous *b*-type transitions analogous to those of $D_2O$-$CO_2$. But we also noticed weaker lines which turned out to be *a*-type transitions, rigorously forbidden in $H_2O$- or $D_2O$- $CO_2$. This means that the HDO-$CO_2$ spectrum needs to be analyzed as a single group of transitions involving four linked states (two tunneling components in each of the ground and excited vibrational states), rather than analyzed as two independent bands as for $H_2O$- and $D_2O$-$CO_2$. The *a*-type transitions are weaker than the *b*-type because the projection of the O-D bond on the *a*-axis is smaller.

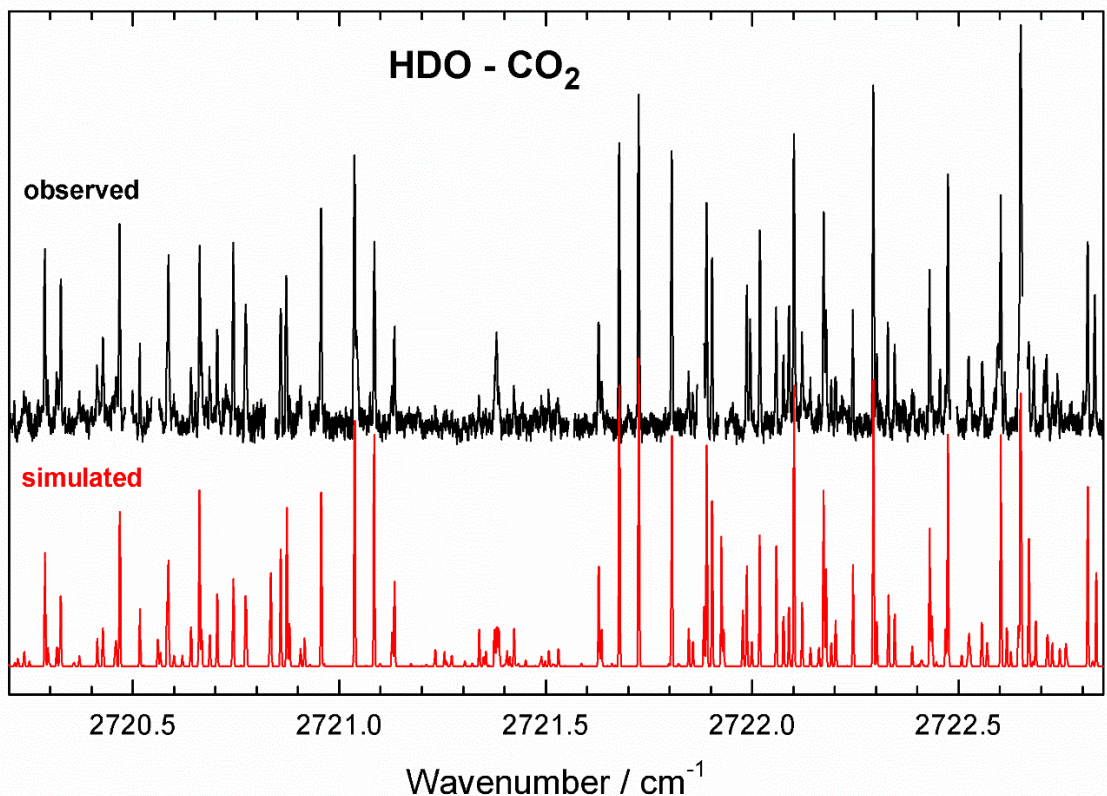

Figure 4. Observed and simulated spectra of HDO-$CO_2$ in the region of the O-D stretch fundamental of HDO.

The spectrum was found to be moderately perturbed (more so than with $D_2O$-$CO_2$ in the $D_2O$ region), so we first analyzed the ground state using the previous microwave data[2,3] plus infrared combination differences. Since we observed both *a*- and *b*-type infrared transitions, this analysis enabled direct determination of the ground state tunneling splitting (a first for water-$CO_2$). These ground state results are given in the lower part of Table 3. As noted also by Columberg et al.,[3] it was necessary to use separate distortion parameter sets for the two tunneling components of HDO-$CO_2$ to obtain a good fit, even though common sets sufficed for $H_2O$- and $D_2O$-$CO_2$.

Table 3. Molecular parameters for $CO_2$-HDO (in $cm^{-1}$). [a]

| | $K_a'$ odd $K_a''$ even | $K_a'$ even $K_a''$ odd |
|---|---|---|
| $\sigma_0'$ | 2721.4051(1) | 2721.3820(2) |
| $A'$ | [0.379] | 0.379464(68) |
| $B'$ | 0.147591(76) | 0.14689(40) |
| $C'$ | 0.105968(93) | 0.10670(38) |
| $\Delta_{JK}'$ | -1.82(40)e-5 | [1.0610e-5] |
| $\Delta_J'$ | -4.3(14)e-6 | 4.5(20)e-6 |
| $\delta_K'$ | -1.30(46)e-4 | -2.5(11)e-4 |
| $\delta_J'$ | [2.6596e-7] | [2.227e-7] |
| $\sigma_0''$ | 0.00 | 0.023443(67) |
| $A''$ | 0.379584(19) | 0.379396(36) |
| $B''$ | 0.14787473(87) | 0.14787264(108) |
| $C''$ | 0.10581199(83) | 0.10580041(93) |
| $\Delta_{JK}''$ | 1.0610(13)e-5 | 8.15(70)e-6 |
| $\Delta_J''$ | 8.847(27)e-7 | 8.263(87)e-7 |
| $\delta_K''$ | 8.705(51)e-6 | 7.54(38)e-6 |
| $\delta_J''$ | 2.227(36)e-7 | 2.6596(93)e-7 |

[a] Quantities in parentheses are $1\sigma$ from the least-squares fit, in units of the last quoted digit. Ground state parameters are from our fit to previous microwave data[2,3] plus selected ground state combination differences from the present spectra. Note that $\sigma_0$ represents vibrational term value energy, not band origin. Distortion parameter $\Delta_K$ could not be determined and was fixed to zero for all states. Excited state distortion parameters in square brackets were fixed at their ground state values for the appropriate (lower or upper) tunneling component. Excited state distortion parameters which were varied may not be meaningful (see text).

We assigned a total of 81 lines in terms of 83 infrared transitions (two blended lines) with values of $K_a' = 0$ to 3 and $J' = 0$ to 8. Of these, 22 were $a$-type and the remainder $b$-type. The spectrum was analyzed with ground state parameters fixed, and the results are given in the upper part of Table 3. The value of the excited state $A$ rotational parameter could not be reliably determined for the upper tunneling component and so was fixed at the value indicated in the table. As mentioned, the HDO-$CO_2$ spectrum was noticeably perturbed. Some transitions involving obviously perturbed upper state levels were given reduced weights, and the overall weighted rms error of the fit was 0.00041 cm$^{-1}$. Perturbed upper state levels, as verified by two or more transitions, included $J_{Ka,Kc} = 4_{2,3}$, with a residual (obs - calc) of about -0.0013 cm$^{-1}$, $6_{1,6}$ (-0.0038 cm$^{-1}$), and $5_{2,4}$ (+0.0205 cm$^{-1}$). There are likely to be other levels with even larger perturbations which were not recognized because their transitions were too shifted from the expected positions.

## 4.    Discussion and conclusions

### 4.1. Tunneling splittings

As noted, in the ground vibrational state of a water-C$^{16}$O$_2$ complex only $K_a$ = even levels are allowed for the lower internal rotation tunneling component and only $K_a$ = odd levels for the upper tunneling component. This is the case whether water = $H_2O$, $D_2O$, or HDO, since it depends on interchange symmetry of the two carbon dioxide $^{16}$O nuclei (note that Columberg et al.[3] appear to miss this point, saying: "*Rotational transitions of HDO-CO$_2$ were observed only in the lowest internal rotation state …*"). For $H_2O$-$CO_2$, H nuclei interchange symmetry means that the lower tunneling component is the *para* spin modification (spin weight 1) and the upper component is the *ortho* modification (spin weight 3). For $D_2O$-$CO_2$, D interchange symmetry means that the lower component is *ortho* (spin weight 6) and the upper component *para* (spin weight 3). For HDO-$CO_2$, there is of course no interchange symmetry between H and D and no *para* / *ortho* distinction.

The transitions observed for infrared bands of water-$CO_2$ depend on the symmetry of the excited vibrational state, as illustrated for $H_2O$-$CO_2$ on the left side of Fig. 5. $D_2O$-$CO_2$ is the same except that the *para* and *ortho* labels are interchanged (the magnitude of the tunneling splittings are also very different). We see in Fig. 5 that the (010) bending vibration of $H_2O$ has $A_1'$ symmetry, same as the ground state, and that the separation between the two infrared bands (*para* and *ortho*) is equal to the difference between the ground and excited state tunneling splittings (observed by Zhu et al.[5] to be 0.007 cm$^{-1}$ for $H_2O$-$CO_2$ and 0.0004 cm$^{-1}$ for $D_2O$-$CO_2$).

The (001) asymmetric stretch vibration of $CO_2$ has $B_2'$ symmetry, so the separation between the infrared bands is again the difference between the ground and excited state splittings (observed here to be 0.0016 cm$^{-1}$ for $H_2O$-$CO_2$ and 0.0002 cm$^{-1}$ for $D_2O$-$CO_2$). The (001) asymmetric stretch vibration of water has $B_2''$ symmetry, so the separation between the infrared bands is now the sum of the ground and excited state splittings (observed by Block et al.[4] to be 0.2681 cm$^{-1}$ for $H_2O$-$CO_2$ and here to be 0.0058 cm$^{-1}$ for $D_2O$-$CO_2$). These band origin separations are summarized in Table 4, which is an expansion of Table 2 of Bogomolov et al.[7], using their notation of $S''$ and $S'$ for the ground and excited tunneling splittings. Symmetries for the $H_2O$ vibrational states (101) and (200) are $B_2''$ and $A_1'$, respectively.

The situation for HDO-$CO_2$ is illustrated on the right hand side of Fig. 5. There is no *para* / *ortho* distinction, so in the O-D stretch region we actually observe four bands with four different band origins, and these enable us to separately determine the ground and excited state tunneling splittings, as summarized in Table 5. The Table also includes data for $H_2O$- and $D_2O$-$CO_2$, where it is not possible to directly determine experimental tunneling splittings, only sums or differences. Fortunately there is good evidence, from the differences in Table 4 and from HDO-$CO_2$, that splittings do not change much with vibrational excitation, at least for intramolecular fundamentals. So it is a good approximation to assume that the ground state splitting is about half of the band origin difference observed for the $H_2O$ or $D_2O$ asymmetric stretch (001) region. Recent theoretical predictions by Felker and Bačić[10] for the $H_2O$- and $D_2O$-$CO_2$ splittings are included in Table 5, and we see that they agree quite well with experiment. It is interesting to note that the HDO-$CO_2$ splitting, 0.02344 cm$^{-1}$, is fairly close to the geometric mean of the $H_2O$- and $D_2O$-$CO_2$ splittings, 0.0198 cm$^{-1}$.

In principle, it should be possible to observe microwave transitions *between* the tunneling states of HDO-$CO_2$ (guided by the present results) which would give an even more precise measurement of the splitting. These would of course be *b*-type transitions, for example $1_{1,1} \leftarrow 0_{0,0}$ which we would predict at about 15249 MHz. However, such transitions are likely to be very weak because of the small projection of the HDO permanent dipole moment on the *b*-axis of HDO-$CO_2$. In a rigid model, the b-type intensity is predicted to be 0.0008 times the a-type one.

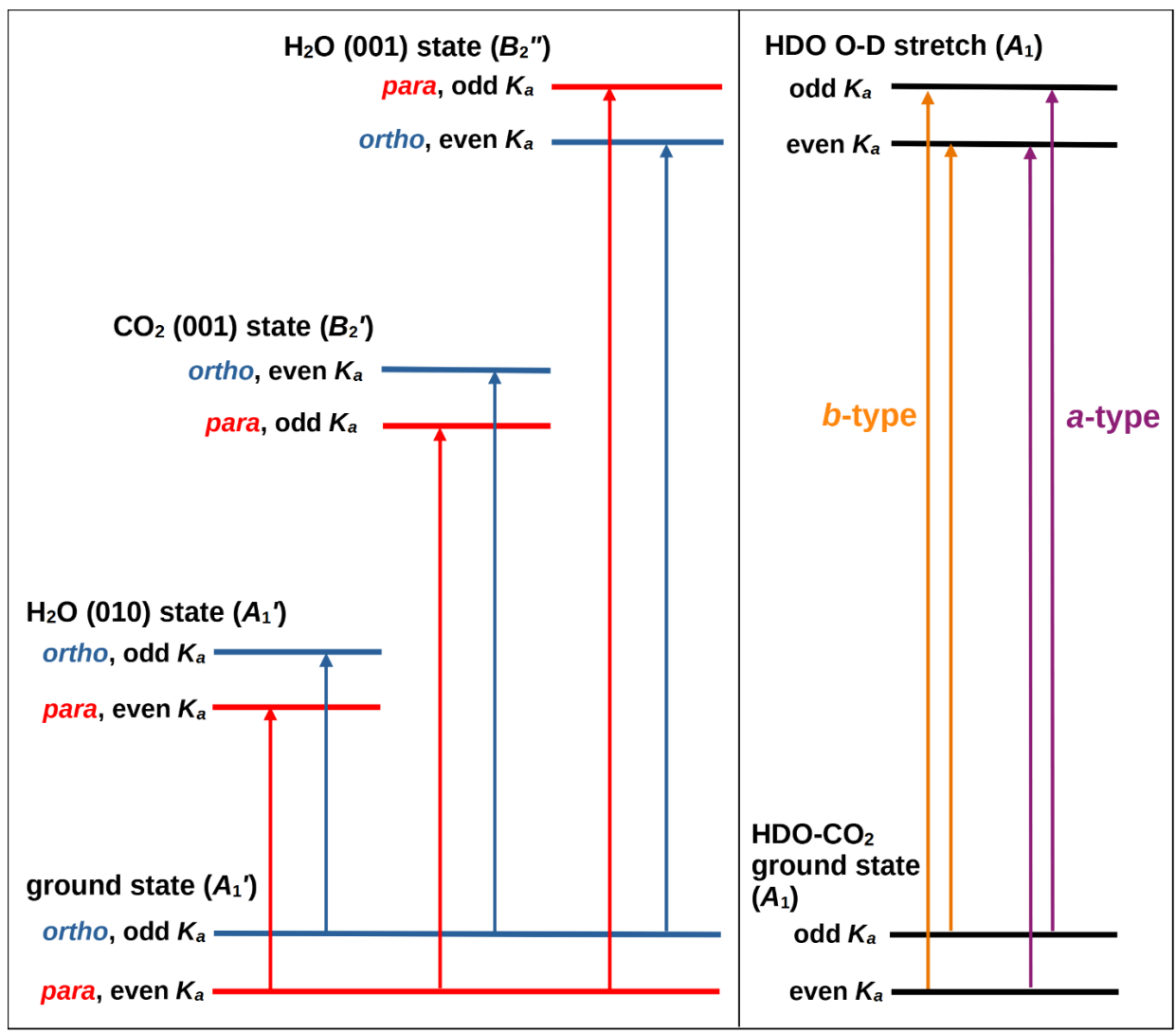

Figure 5. Illustration of allowed infrared transitions among internal rotation tunneling components for different infrared bands of $H_2O$-$CO_2$ (left) and HDO-$CO_2$ (right). $D_2O$-$CO_2$ is just the same as $H_2O$-$CO_2$, except that the *para* and *ortho* labels are interchanged. Vibrational symmetry labels ($A_1'$, etc.) refer to the $G_8$ (MS) group for $H_2O$-$CO_2$ or the $G_4$ (MS) group for HDO-$CO_2$.

Table 4. Observed band origin separations, $\Delta\nu_0$, for infrared bands of water - $CO_2$ complexes (in $cm^{-1}$).[a]

| | upper state | $\Delta\nu_0$ | splitting sum or difference | reference |
|---|---|---|---|---|
| $H_2O$-$CO_2$ | $H_2O$ (010) | +0.0070 | $S'' - S'$ | Zhu et al.[5] |
| | $CO_2$ (001) | -0.0016 | $S'' - S'$ | present work |
| | $H_2O$ (001) | 0.2681 | $S'' + S'$ | Block et al.[4] |
| | $H_2O$ (200) | +0.0168 | $S'' - S'$ | Bogomolov et al.[7] |
| | $H_2O$ (101) | 0.1948 | $S'' + S'$ | Bogomolov et al.[7] |
| $D_2O$-$CO_2$ | $D_2O$ (010) | +0.0004 | $S'' - S'$ | Zhu et al.[5] |
| | $CO_2$ (001) | +0.0002 | $S'' - S'$ | present work |
| | $D_2O$ (001) | 0.0058 | $S'' + S'$ | present work |

[a] Depending on the upper state vibrational symmetry, the band origin separation is equal either to the sum or the difference of the ground (S") and excited (S') state tunneling splittings.

Table 5. Ground ($S''$) and excited ($S'$) state tunneling splittings for water - $CO_2$ complexes from the fundamental O-H or O-D stretching region (in $cm^{-1}$).

| | $S'' + S'$ | $S''$ | $S'$ | $S''$ calculated [a] |
|---|---|---|---|---|
| $H_2O$-$CO_2$ [b] | 0.2681 | 0.1356 | 0.1325 | 0.1447 |
| HDO-$CO_2$ [c] | 0.0465 | 0.02344 | 0.0231 | |
| $D_2O$-$CO_2$ [d] | 0.0058 | 0.0029 | 0.0029 | 0.0039 |

[a] Felker and Bačić.[10]

[b] Block et al.[4] We are not sure exactly how the separate values of $S''$ and $S'$ were determined for $H_2O$-$CO_2$ in Ref. 4. Simply assuming equal ground and excited state splittings gives $S'' = S' = 0.134$ $cm^{-1}$.

[c] Present work. Tunneling splittings $S''$ and $S'$ for HDO-$CO_2$ were determined independently for the ground and excited states, which is not possible for $H_2O$- and $D_2O$-$CO_2$.

[d] Present work. We assume $S'' = S'$ for $D_2O$-$CO_2$.

### 4.2. Vibrational shifts

Table 6 summarizes observed vibrational shifts for water-$CO_2$ complexes (with tunneling components averaged) relative to the vibrational frequencies of the free water or $CO_2$ molecules. There are small positive (blue) shifts for the water bend and $CO_2$ asymmetric stretch fundamentals, and negative (red) shifts for all the observed O-H and O-D stretch vibrations. Interestingly, the HDO-$CO_2$ shift in the O-D stretch region is almost exactly midway between those of $H_2O$- and $D_2O$-$CO_2$ in their (001) regions, even though the nature of the HDO vibration is somewhat different (since in a sense the O-D stretch in HDO is a mixture of the $H_2O$ and $D_2O$ (100) and (001) modes). Also listed in Table 6 are theoretical harmonic vibrational shifts as calculated by Makarewicz,[8] which are all somewhat larger in magnitude than the observed ones.

Table 6. Vibrational shifts for water - $CO_2$ complexes (in $cm^{-1}$).

|  | water (010) | $CO_2$ (001) | water (001) | water (200) | water (101) |
|---|---|---|---|---|---|
| Theory [a] |  |  |  |  |  |
| $H_2O$-$CO_2$ | -2.2 | +3.0 | -3.2 |  |  |
| Experiment |  |  |  |  |  |
| $H_2O$-$CO_2$ | +0.739 [b] | +1.159 [c] | -3.020 [e] | -6.134 [f] | -2.824 [f] |
| HDO-$CO_2$ |  |  | -2.282 [c,d] |  |  |
| $D_2O$-$CO_2$ | +0.083 [b] | +1.169 [c] | -1.485 [c] |  |  |

[a] Makarewicz.[8]

[b] Zhu et al.[5]

[c] Present results.

[d] This vibration is the O-D stretch of HDO.

[e] Block et al.[4]

[f] Bogomolov et al.[7]

The isotope dependence of the shifts is small, or at least regular, except for the water (010) bending region, where the $H_2O$-$CO_2$ shift is much larger in magnitude than that of $D_2O$-$CO_2$.[5] This is also the only region where the observed and calculated shifts disagree in sign.

Could these differences be related to vibrational mixing and the fact that the $D_2O$-$CO_2$ vibration (1178 cm$^{-1}$) lies below the (020) and (100) vibrations of $CO_2$ (1285 and 1388 cm$^{-1}$), while the $H_2O$-$CO_2$ vibration (1595 cm$^{-1}$) lies well above?

### 4.3. Conclusions

In conclusion, four new infrared bands of the water-$CO_2$ dimer have been observed and analyzed: $H_2O$-$CO_2$ and $D_2O$-$CO_2$ in the $CO_2$ $\nu_3$ region ($\approx$2350 cm$^{-1}$), $D_2O$-$CO_2$ in the $D_2O$ $\nu_3$ region ($\approx$2790 cm$^{-1}$), and HDO-$CO_2$ ($\approx$2720 cm$^{-1}$). These represent the first observation of water-$CO_2$ spectra accompanying a $CO_2$ vibration, and the first infrared observation of HDO-$CO_2$. As a result, we obtain the first experimental values for the internal rotation tunneling splittings of $D_2O$-$CO_2$ (0.003 cm$^{-1}$) and HDO-$CO_2$ (0.0234 cm$^{-1}$), the latter being a true independent ground state determination thanks to the reduced symmetry of HDO-$CO_2$.

**Acknowledgements**

The financial support of the Natural Sciences and Engineering Research Council of Canada is gratefully acknowledged. C.L. thanks the financial support of the "Communauté Française de Belgique" (ARC IBEAM 18/23-090) and of the FNRS CDR J.0113.23 and J.0129.20.

**References**

[1] Q. Wang and J. M. Bowman, Two-component, *ab initio* potential energy surface for $CO_2$-$H_2O$, extension to the hydrate clathrate, $CO_2$@$(H_2O)_{20}$, and VSCF/VCI vibrational analyses of both, J. Chem. Phys. **147**, 161714 (2017).

[2] K. I. Peterson and W. Klemperer, Structure and internal rotation of $H_2O$-$CO_2$, HDO-$CO_2$, and $D_2O$-$CO_2$ van der Waals complexes, J. Chem. Phys. **80**, 2439-2445 (1984).

[3] G. Columberg, A. Bauder, N. Heineking, W. Stahl, and J. Makarewicz, Internal rotation effects and hyperfine structure in the rotational spectrum of a water-carbon dioxide complex, Mol. Phys. **93**, 215-228 (1998).

[4] P. A. Block, M. D. Marshall, and R. E. Miller, Wide amplitude motion in the water–carbon dioxide and water–acetylene complexes, J. Chem. Phys. **96**, 7321-7332 (1992).

[5] Y. Zhu, S. Li, P. Sun, and C. Duan, Infrared diode laser spectroscopy of $H_2O$-$CO_2$ and $D_2O$-$CO_2$ complexes in the $\nu_2$ bend region of water, J. Mol. Spectrosc. **283**, 7-9 (2013).

[6] C. Lauzin, A. C. Imbreckh, T. Foldes, T. Vanfleteren, N. Moazzen-Ahmadi, and M. Herman, High-resolution spectroscopic study of the $H_2O$–$CO_2$ van der Waals complex in the 2OH overtone range, Mol. Phys. **128**, e1706776 (2020).

[7] A. S. Bogomolov, A. Roucou, R. Bejjani, M. Herman, N. Moazzen-Ahmadi, and C. Lauzin, The rotationally resolved symmetric 2OH excitation in $H_2O$-$CO_2$ observed using pulsed supersonic expansion and CW-CRDS, Chem. Phys. Lett. **774**, 138606 (2021).

[8] J. Makarewicz, Intermolecular potential energy surface of the water-carbon dioxide complex, J. Chem. Phys. **132**, 234305 (2010).

[9] R. J. Wheatley and A. H. Harvey, Intermolecular potential energy surface and second virial coefficients for the water–$CO_2$ dimer, J. Chem. Phys. **134**, 134309 (2011).

[10] P. M. Felker and Z. Bačić, Intermolecular rovibrational states of the $H_2O$–$CO_2$ and $D_2O$–$CO_2$ van der Waals complexes, J. Chem. Phys. **156**, 064301 (2022).

[11] N. Moazzen-Ahmadi and A. R. W. McKellar, Spectroscopy of dimers, trimers, and larger clusters of linear molecules, Int. Rev. Phys. Chem. **32**, 611-650 (2013).

[12] A.J. Barclay, A.R.W. McKellar, and N. Moazzen-Ahmadi, Spectra of the $D_2O$ dimer in the O-D fundamental stretch region: vibrational dependence of tunneling splittings and lifetimes, J. Chem. Phys. **150**, 164307 (2019).

[13] C. M. Western, PGOPHER, a program for simulating rotational structure version 8.0, 2014, University of Bristol Research Data Repository, doi:10.5523/bris.huflggvpcuc1zvliqed497r2

[14] P.R. Bunker and P. Jensen, Molecular symmetry and Spectroscopy, 2nd Edition, NRC Research Press, Ottawa, 2006.